\documentclass[aps, prx, 10pt, amssymb,amsmath,superscriptaddress,tightenlines,twocolumn,notitlepage] {revtex4-2}
\usepackage{graphicx}
\usepackage{amsmath}
\usepackage{amssymb}
\usepackage{bm}
\usepackage{color}
\usepackage{hyperref}
\usepackage{tabularx}
\usepackage{multirow}
\usepackage{booktabs}
\usepackage{braket}
\usepackage{epstopdf}

\makeatletter
\DeclareRobustCommand{\r}{\ifmmode\expandafter\mathring\else\expandafter\@ring\fi}
\makeatother

\begin{document}

\title{Interference-Enhanced Large Electron--Phonon Coupling from Raman-active Breathing Modes in Moir\'e Semiconductors}

\author{Ning Mao}
\thanks{These authors contributed equally to this work.}
\affiliation{Max Planck Institute for Chemical Physics of Solids, 01187, Dresden, Germany}

\author{Shaozheng Wang}
\thanks{These authors contributed equally to this work.}
\affiliation{State Key Laboratory of Micro-nano Engineering Science, Key Laboratory of Artificial Structures and Quantum Control (Ministry of Education), School of Physics and Astronomy, Shanghai Jiao Tong University, Shanghai, China.} 

\author{Cheng Xu}
\affiliation{Max Planck Institute for Chemical Physics of Solids, 01187, Dresden, Germany}

\author{Xumin Chang}
\affiliation{State Key Laboratory of Micro-nano Engineering Science, Key Laboratory of Artificial Structures and Quantum Control (Ministry of Education), School of Physics and Astronomy, Shanghai Jiao Tong University, Shanghai, China.} 

\author{Kenji Watanabe}
\affiliation{National Institute for Materials Science, 1-1 Namiki, Tsukuba 305-0044, Japan}

\author{Takashi Taniguchi}
\affiliation{National Institute for Materials Science, 1-1 Namiki, Tsukuba 305-0044, Japan}

\author{Claudia Felser}
\affiliation{Max Planck Institute for Chemical Physics of Solids, 01187, Dresden, Germany}

\author{Shengwei Jiang}
\email{swjiang@sjtu.edu.cn}
\affiliation{State Key Laboratory of Micro-nano Engineering Science, Key Laboratory of Artificial Structures and Quantum Control (Ministry of Education), School of Physics and Astronomy, Shanghai Jiao Tong University, Shanghai, China.} 

\author{Yang Zhang}
\email{yangzhang@utk.edu}
\affiliation{Department of Physics and Astronomy, University of Tennessee, Knoxville, TN 37996, USA}
\affiliation{Min H. Kao Department of Electrical Engineering and Computer Science, University of Tennessee, Knoxville, Tennessee 37996, USA}

\begin{abstract}
Superconductivity was recently observed in twisted WSe$_2$ and MoTe$_2$, raising a central question: is the pairing driven by electronic correlations, by phonons, or by both? Answering it requires determining the electron--phonon coupling (EPC) in these moir\'e semiconductors, whose calculation in realistic supercells of thousands of atoms lies beyond the reach of direct first-principles methods. Here we combine filling-dependent Raman spectroscopy with machine-learning first-principles calculations to obtain the EPC mode by mode in supercells of up to tens of thousands of atoms. Raman reveals only a few moir\'e phonons whose frequencies shift strongly with filling; we trace this to an interference selection rule: a phonon couples strongly only when its displacement texture matches the static lattice-reconstruction pattern, and is otherwise suppressed by destructive interference. The rule selects the low- and high-frequency breathing modes seen in Raman and makes the coupling peak at large twist angles, near those at which superconductivity appears. Lattice-reconstruction interference thus emerges as an organizing principle for moir\'e EPC, pointing to a substantial, potentially dominant, phonon contribution to large-angle pairing.
\end{abstract}
\maketitle


\section{Introduction}
Moir\'e superlattices of transition metal dichalcogenides (TMDs) have emerged as a highly tunable platform for correlated and topological electronic phases, hosting flat bands whose bandwidth, filling, and topology can be controlled continuously by the twist angle and by electrostatic gating.
Within this platform, superconductivity has recently been observed in twisted bilayer WSe$_2$ (t-WSe$_2$) and twisted bilayer MoTe$_2$ (t-MoTe$_2$), with critical temperatures on the order of hundreds of millikelvin near integer filling ($\nu=-1$)~\cite{xia2025superconductivity,guo2025superconductivity,guo2025angle,xia2025simulating,xu2025signatures,sun2026twist}, and its microscopic pairing mechanism is now under active debate.
A widely discussed scenario attributes the pairing primarily to repulsive electronic correlations, such as the Kohn--Luttinger mechanism, which generates an effective attraction in specific angular-momentum channels despite the bare Coulomb repulsion~\cite{schrade2024nematic,tuo2025theory,qin2024kohn,chubukov2025quantum,zhu2025plane,fischer2025theory}.
However, correlation-based pairing does not preclude an important phonon-mediated contribution~\cite{wu2018theory}, especially near $\theta\approx 5^{\circ}$, where superconductivity emerges from a metallic background far from any Mott-insulating state.
Moir\'e systems can enhance electron--phonon coupling (EPC) through an increased density of states (DOS) from flat bands and van Hove singularities, through twist-tunable band structures, and through low-energy lattice modes~\cite{cai2023signatures,zeng2023thermodynamic,park2023observation,xu2023observation,xu2025signatures,christos2025approximate,zhu2025superconductivity,xie2025superconductivity}.
The open question is whether the EPC in twisted TMDs is strong enough to drive the pairing, rather than merely assist it.

A major obstacle to answering this question is that realistic moir\'e unit cells at small twist angles contain thousands to tens of thousands of atoms, which makes direct first-principles calculations of lattice relaxation, phonons, and EPC prohibitively expensive. As a result, previous studies have largely relied on continuum elasticity theories or classical interatomic potentials~\cite{lu2022low,liu2022moire,ramos2025flat,maity2020phonons,krisna2023moire,pizzi2021shear,lin2018moire,bathen2025precise,koshino2019moir,xie2023lattice,zhu2025superconductivity,xie2025superconductivity}, which often neglect the intricate lattice-reconstruction patterns and specific phonon-mode characters that are central to the moir\'e electronic structure.
Machine-learning methods have recently brought EPC calculations to near-DFT accuracy in simple systems such as GaAs and large-angle twisted bilayer graphene, where the unit cell contains only a few hundred atoms~\cite{li2024deep,zhong2024accelerating}.
Extending this capability to moir\'e TMDs is far more demanding: strong spin-orbit coupling, multi-orbital interlayer hybridization, and substantial lattice reconstruction all enter the problem simultaneously, while the commensurate unit cell grows to thousands of atoms. A quantitative first-principles description of moir\'e phonons and their coupling to the electronic degrees of freedom is therefore still lacking.

In this work, we combine filling-dependent Raman spectroscopy with large-scale first-principles simulation to resolve the EPC in twisted TMDs mode by mode. We first perform Raman measurements on t-WSe$_2$ devices with twist angles of $4.4^\circ$ and $3.15^\circ$. Although the number of moir\'e phonon modes grows drastically as the twist angle decreases, both devices exhibit only three Raman peaks, whose positions shift systematically with hole filling and reveal strong coupling of these modes to the carriers. To account for these observations, we develop a scalable machine-learning framework for moir\'e phonons and their EPC (Fig.~S1) and apply it to t-WSe$_2$ and t-MoTe$_2$ across twist angles from $1.89^\circ$ to $13^\circ$.
We discover an interference mechanism governing the EPC in twisted TMDs: phonon modes whose displacement texture matches the static lattice-reconstruction pattern couple strongly to the electronic structure, whereas modes with mismatched textures are suppressed. We quantify this correspondence with a pattern-matching parameter $\chi$ that projects each phonon eigenvector onto the reconstruction displacement field; the breathing mode in Fig.~\ref{moire-phonon}a then exhibits strong EPC ($\chi \approx 1$), whereas the shear mode in Fig.~\ref{moire-phonon}b couples negligibly ($\chi \approx 0$).
As a direct consequence, the total coupling constant $\lambda$ peaks at a large twist angle of $7.34^\circ$ for both t-WSe$_2$ and t-MoTe$_2$, in contrast to the density of states, which increases toward smaller angles. Because this places the strongest coupling in the regime where superconductivity is observed~\cite{guo2025angle}, it points to a leading role for phonons in large-angle pairing.

\begin{figure*}[t]
\includegraphics[width=2\columnwidth]{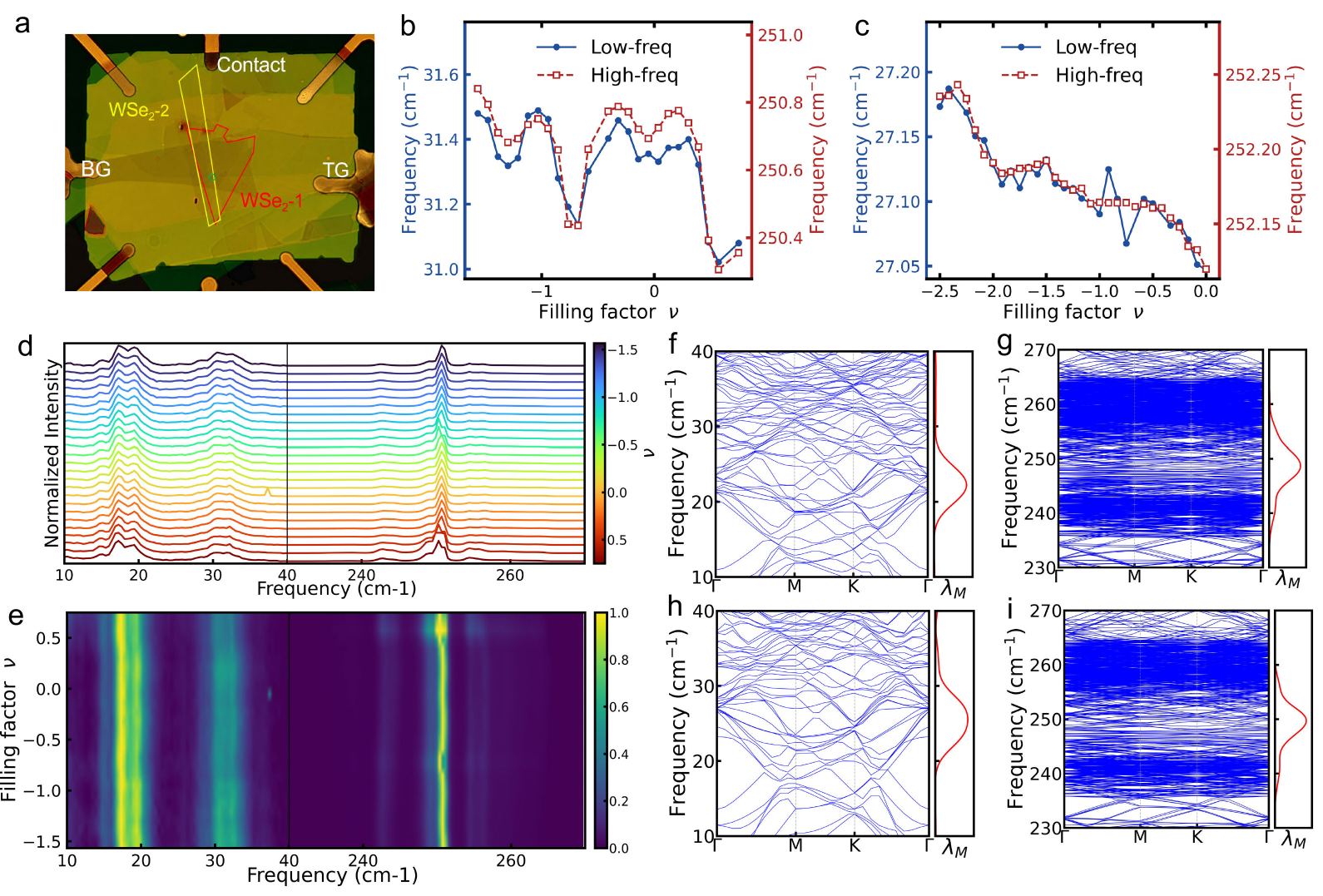}
\caption{
\textbf{Filling-dependent Raman spectroscopy of moir\'e phonons in t-WSe$_2$.}
\textbf{a}, Device used for Raman characterization of t-WSe$_2$ at $4.4^\circ$. The red and yellow regions correspond to the two individual layers, and the green dots mark the positions of the Raman laser spot.
\textbf{b}, Peak positions of the low- and high-frequency modes as a function of filling factor $\nu$ for t-WSe$_2$ at $4.4^\circ$.
\textbf{c}, Peak positions of the low- and high-frequency modes as a function of filling factor $\nu$ for t-WSe$_2$ at $3.15^\circ$.
\textbf{d}, Stacked Raman spectra at different filling factors $\nu$, vertically offset for clarity.
\textbf{e}, False-color map of the normalized Raman intensity as a function of frequency and filling factor $\nu$.
\textbf{f}, \textbf{g}, Phonon dispersion (left) and mode-resolved electron--phonon coupling (EPC) strength (right) for t-WSe$_2$ at $4.4^\circ$, shown for (\textbf{f}) the low-frequency and (\textbf{g}) the high-frequency phonon regions.
\textbf{h}, \textbf{i}, Phonon dispersion (left) and mode-resolved electron--phonon coupling (EPC) strength (right) for t-WSe$_2$ at $5.08^\circ$, shown for (\textbf{h}) the low-frequency and (\textbf{i}) the high-frequency phonon regions.
}
\label{raman}
\end{figure*}

\section{Filling-dependent Raman signatures}

At the twist angles of interest, a moir\'e supercell of t-WSe$_2$ hosts thousands of phonon branches, yet only a symmetry-selected few are Raman-active. Whether those few are sensitive to the carriers, and hence to the pairing, is the question our measurements address.

We perform low-temperature ($1.7$~K) confocal Raman spectroscopy on dual-gated t-WSe$_2$ devices at twist angles of $4.4^\circ$ and $3.15^\circ$, the former within the range where superconductivity is observed~\cite{xia2025superconductivity,guo2025superconductivity,guo2025angle,xia2025simulating}; the devices are shown in Figs.~\ref{raman}a and~S14. Sweeping the hole filling $\nu$ through the first moir\'e band by electrostatic gating, we find that each device exhibits only three Raman-active moir\'e phonons despite the proliferation of folded branches: two in the low-frequency interlayer region, at $18$ and $31~\mathrm{cm}^{-1}$, and one at $250~\mathrm{cm}^{-1}$ (Figs.~\ref{raman}d,e). Comparison with the calculated spectra (Figs.~\ref{raman}f--i) identifies the low-frequency pair as the layer breathing mode (LBM) and the layer shear mode (LSM), and the high-frequency mode as an optical breathing mode (OBM).

The central experimental observation is that these modes are markedly gate-tunable. As $\nu$ is swept, their peak positions shift systematically (Figs.~\ref{raman}b,c) by up to $\sim$0.1~meV, a large excursion for a phonon energy and direct evidence that the Raman-active modes couple strongly to the carriers. The LBM and OBM exhibit closely similar filling dependences, with the OBM showing a larger frequency shift. The response is, moreover, sharply twist-angle dependent: the total peak excursion reaches $0.6~\mathrm{cm}^{-1}$ at $4.4^\circ$ but only $0.15~\mathrm{cm}^{-1}$ at $3.15^\circ$. Remarkably, the coupling thus grows with twist angle, opposite to the density of states, which increases toward smaller angles.

Two features stand out: the electron--phonon coupling in t-WSe$_2$ is strong and highly mode-selective, with only a handful of the many moir\'e phonons responding to the carriers, and it is largest at the larger twist angle. Filling-induced frequency shifts do not by themselves quantify the coupling, and existing continuum or classical-potential treatments capture only the low-frequency interlayer modes such as the LSM and LBM and leave the higher-frequency moir\'e phonons unexplored~\cite{lu2022low,liu2022moire,ramos2025flat,maity2020phonons,krisna2023moire,pizzi2021shear,lin2018moire,bathen2025precise,koshino2019moir,xie2023lattice,zhu2025superconductivity,xie2025superconductivity}. Explaining which moir\'e phonons couple, and why the coupling strengthens with angle, requires a quantitative first-principles account of moir\'e phonons and their EPC.

\section{Lattice reconstruction and moir\'e phonons}
\subsection{Lattice relaxation}
\begin{figure*}[t]
\includegraphics[width=2\columnwidth]{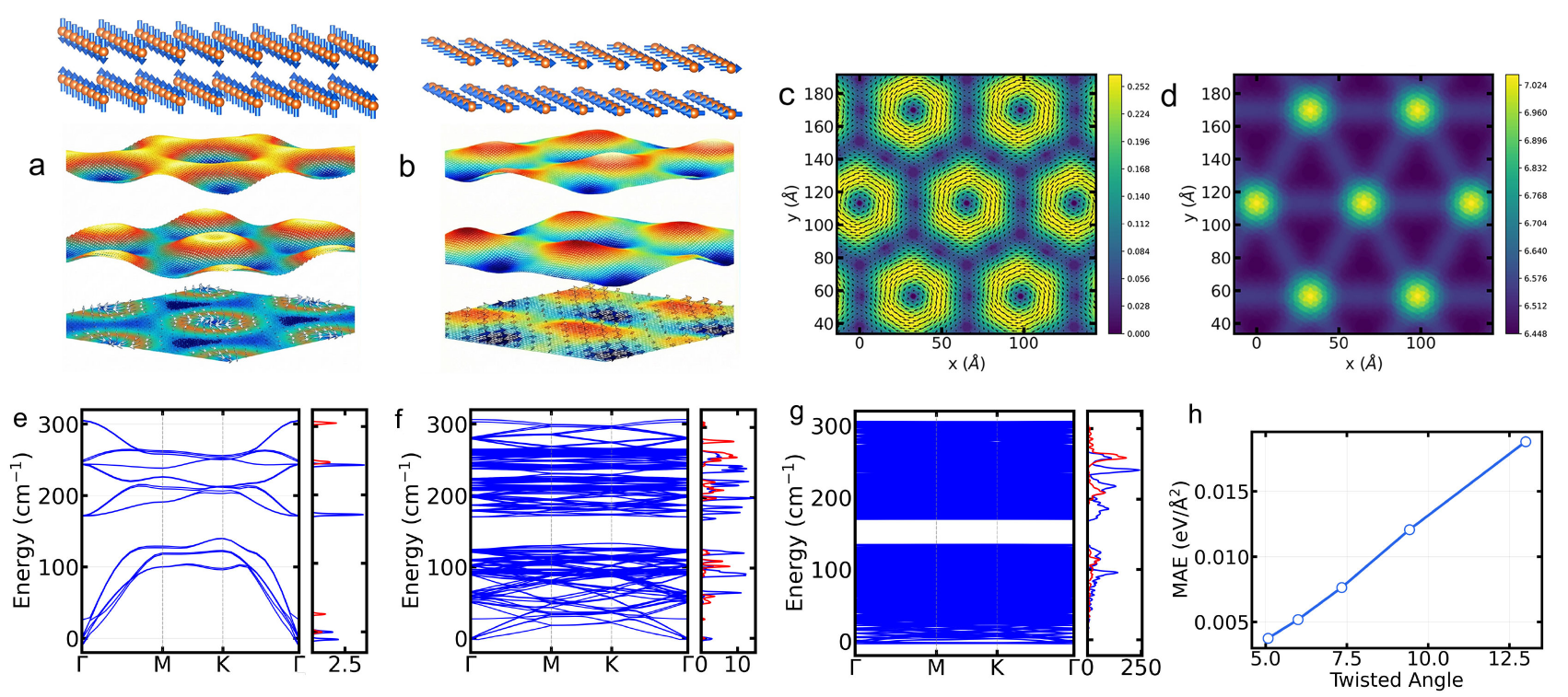}
\caption{ 
\textbf{Lattice relaxation in twisted TMDs and its impact on the moir\'e phonon spectrum.}
\textbf{a}, \textbf{b}, Schematic illustrations of the (\textbf{a}) layer breathing mode (LBM) and (\textbf{b}) layer shear mode (LSM); in each, the upper panel shows the mode in the bilayer and the lower panel the corresponding moir\'e phonon.
\textbf{c}, \textbf{d}, Intralayer displacement (\textbf{c}) and interlayer distance (\textbf{d}) maps of the relaxed $2.88^\circ$ t-WSe$_2$ structure obtained from density functional theory. The former shows the in-plane relaxation-induced displacements of W atoms in the top layer, and the latter the vertical separation between W atoms in the top and bottom layers.
\textbf{e}--\textbf{g}, Phonon dispersions of (\textbf{e}) untwisted bilayer WSe$_2$ and t-WSe$_2$ at twist angles of (\textbf{f}) $13^\circ$ and (\textbf{g}) $1.89^\circ$. The right subplots show the phonon DOS at the $\Gamma$ point, where blue and red lines represent the in-plane and out-of-plane vibrational contributions. 
\textbf{h}, Mean absolute error of force constants between MLFF model predictions and first-principles calculations. 
}
\label{moire-phonon}
\end{figure*}

In moir\'e superlattices, relaxation is a collective reconstruction of the atomic structure that reshapes the electronic properties. A correctly relaxed structure, in which the residual atomic forces are minimized, is also a prerequisite for the phonon calculation: an insufficiently relaxed structure produces imaginary frequencies in the phonon spectrum.
Using machine-learning force fields (MLFFs), we obtain fully relaxed structures for twist angles from $13^\circ$ down to $1.89^\circ$~\cite{mao2024transfer}.
In these structures, lattice relaxation is described by an in-plane displacement field $\mathbf{u}(\mathbf{r})$ together with an out-of-plane modulation $h(\mathbf{r})$ of the interlayer distance, which jointly reshape the local stackings (MM, MX, XM) into well-defined moir\'e domains~\cite{jia2024moire,mao2024transfer,zhang2024polarization,liu2025dpmoire,qi2026chern,yang2024evolution}.
The in-plane reconstruction arises from the competition between the intralayer elastic energy and interlayer stacking energy.
As shown in Figs.~S2--S3, decreasing the twist angle progressively shrinks the high-energy MM regions while expanding the low-energy MX/XM domains.
These domains are separated by a domain-wall network that concentrates most of the strain, transforming the nearly sinusoidal moir\'e modulation of the rigid lattice into an effective landscape of wells and barriers.
As further shown in Figs.~\ref{moire-phonon}c and~\ref{moire-phonon}d, the in-plane relaxation takes the form of a collective, vortex-like rotation of atoms around the MM region. 
The out-of-plane relaxation produces a pronounced corrugation of the interlayer separation. The MM regions exhibit the largest spacing ($\approx7~\mathrm{\AA}$), whereas the MX/XM regions exhibit the smallest ($\approx6.4~\mathrm{\AA}$), giving a corrugation amplitude of about 0.6~$\mathrm{\AA}$.
This interlayer-distance texture is locked to the in-plane reconstruction and generates an overall honeycomb-like domain pattern, in which MM regions form the nodes and MX/XM regions form the connected network.



\subsection{Phonon spectrum}The phonon frequencies and eigenvectors of the relaxed structures follow from the harmonic interatomic force constants (IFCs), the second derivatives of the total energy with respect to atomic displacements. Because a direct first-principles evaluation of the IFCs over all displacements is prohibitive for moir\'e cells, we compute the restoring forces with the same MLFF used for the relaxation and build the IFCs by finite differences. We then Fourier-transform these into the dynamical matrix and diagonalize it to obtain the phonon branches (see Methods). Benchmarked against first-principles IFCs for twist angles from $13^\circ$ down to $5.08^\circ$, the mean absolute error falls below $0.005~\mathrm{eV/\AA^2}$ at $5.08^\circ$ [Fig.~\ref{moire-phonon}h], so the MLFF reproduces the force constants with near-DFT accuracy.

As shown in Fig.~\ref{moire-phonon}e, the spectrum of bilayer WSe$_2$ contains six phonon branches with frequencies below 150 cm$^{-1}$. All six reach their minimum at the $\Gamma$ point, where their frequencies drop below 30 cm$^{-1}$. Among these, three are acoustic modes, corresponding to rigid translations along the $x$, $y$, and $z$ axes. Two are layer shear modes (LSMs), involving out-of-phase in-plane vibrations of the constituent layers. The remaining one is the layer breathing mode (LBM) with a frequency of 26 cm$^{-1}$, characterized by out-of-phase vibrations along the $z$-axis.

We now follow these six branches into the moir\'e superlattice, where the crystal symmetry is lowered from that of the pristine bilayer to space group $P321$ (No. 150) for the R-type stacking. Despite this symmetry reduction, the six low-frequency phonon branches of the pristine bilayer remain identifiable in the moir\'e system.
As depicted in Fig.~\ref{moire-phonon}f for a twist angle of $13^\circ$, a flat phonon branch originating from the pristine LBM remains isolated at the $\Gamma$ point with a frequency of $\sim$20 cm$^{-1}$. At the same time, phonon folding from finite $\mathbf{q}$-points of the pristine Brillouin zone into the $\Gamma$ point of the moir\'e Brillouin zone (mBZ) mixes the folded modes with the original phonons. This hybridization reconstructs the original vibrations into distinct moir\'e phonons; the branches appearing immediately above the LBM in Fig.~\ref{moire-phonon}f arise from such mixing. As the twist angle decreases, the number of folded phonons grows rapidly [Figs.~S4 and~S5], and the spectrum becomes densely populated, as shown in Fig.~\ref{moire-phonon}g for $\theta = 1.89^\circ$.

\subsection{Breathing and shear modes}While the eigenvalues are largely inherited from the bilayer phonons, the eigenvectors carry the distinctive signature of the moir\'e reconstruction.
A prominent example is the LBM, which characterizes the relative vibration of the two layers along the out-of-plane $z$-direction. In pristine bilayers, the LBM manifests as a rigid, out-of-phase oscillation where the upper and lower layers vibrate in opposite directions along the $z$-axis, as illustrated in Fig.~S6. 
However, in twisted systems, the moir\'e potential induces spatial variations in the equilibrium interlayer spacing across the unit cell.
This reconstruction affects both the out-of-plane and in-plane components of the LBM. For the out-of-plane component, the interlayer spacing is maximized in MM regions and minimized in MX/XM regions [Figs.~\ref{vibration-pattern}a--\ref{vibration-pattern}b], so the vibration amplitude is largest in the MM regions.
As the twist angle decreases, the LBM frequency softens and the mode becomes more spatially localized. Unlike the pristine case, where the LBM is strictly out-of-plane, the twisted bilayer develops a sizable in-plane component that grows as the twist angle decreases.
These appear as in-plane chiral modes, in which vibrations of distinct chirality couple to specific layers, with the chiral centers aligned to the MM regions.


The other special low-frequency mode is the LSM, the relative in-plane sliding of the two layers. Despite its low frequency, it is distinct from the acoustic branches, separated by a finite frequency gap at the $\Gamma$ point. Its $x$ and $y$ components are degenerate at $\Gamma$ and purely in-plane. In the twisted bilayer, local variations in stacking order break the global sliding symmetry, so the LSM eigenvector is no longer aligned along a single axis but mixes different in-plane directions, increasingly so as the twist angle decreases [Figs.~\ref{vibration-pattern}d,e].
At a twist angle of $5.08^{\circ}$, the LSM exhibits in-plane vibrations predominantly aligned with the $x$-axis, despite showing some local deviations in certain regions. In contrast, at a twist angle of $1.89^{\circ}$, the LSM displays simultaneous in-plane vibrations along three $C_3$-symmetric directions.
These vibrations localize within soliton-like domain walls, with negligible amplitude inside the domains, marking a transition of the LSM from the rigid sliding of the bilayer to a collective vibration of the domain-wall network. Because of its very low frequency, the LSM stays decoupled from the out-of-plane mode: its out-of-plane component remains below $5\%$ across the entire range from $22^{\circ}$ down to $1.89^{\circ}$, preserving its in-plane character.

\section{Electron--phonon coupling}
\subsection{Coupling of the low-frequency modes}To determine which moir\'e phonons couple to the electrons, we use the frozen-phonon method, imposing the displacement pattern of each phonon eigenvector onto the equilibrium geometry to generate the distorted configurations
\begin{equation}
\mathbf{R}_{\kappa\alpha}(\pm Q)=\mathbf{R}_{\kappa\alpha}^{0} \pm \frac{Q}{\sqrt{M_\kappa}}\,e^{j}_{\kappa\alpha},
\end{equation}
and computing the resulting band-structure changes with machine-learning Hamiltonians~\cite{bao2025transfer}, as detailed in the Supplementary Material.

\begin{figure*}[t]
\includegraphics[width=1.5\columnwidth]{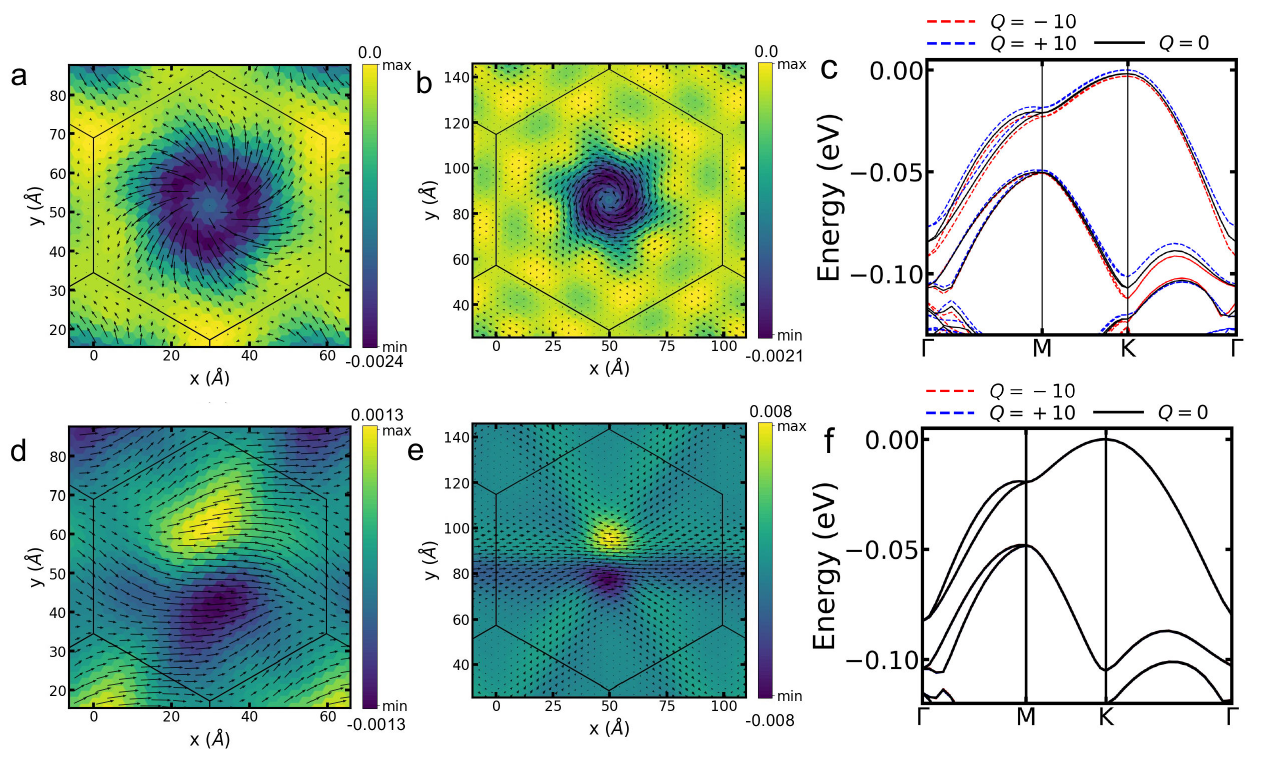}
\caption{ 
\textbf{Influence of the LBM and LSM on the moir\'e electronic structure.}
\textbf{a}, \textbf{b}, Vibration patterns of low-frequency phonon modes in AA-stacked t-WSe$_2$. LBM patterns at twist angles of (\textbf{a}) $5.08^\circ$ and (\textbf{b}) $1.89^\circ$, respectively. 
\textbf{c}, Electronic band structures of t-WSe$_2$ under different LBM distortions.
\textbf{d}, \textbf{e}, LSM patterns at twist angles of (\textbf{d}) $5.08^\circ$ and (\textbf{e}) $1.89^\circ$, respectively. The color map represents the out-of-plane displacement component, while the arrows indicate the in-plane displacement vectors.
\textbf{f}, Electronic band structures of t-WSe$_2$ under different LSM distortions. 
The blue, red, and green lines represent the equilibrium band structure and the band structures under positive and negative distortions, respectively.
}\label{vibration-pattern}
\end{figure*}

\begin{figure*}[t]
\includegraphics[width=1.5\columnwidth]{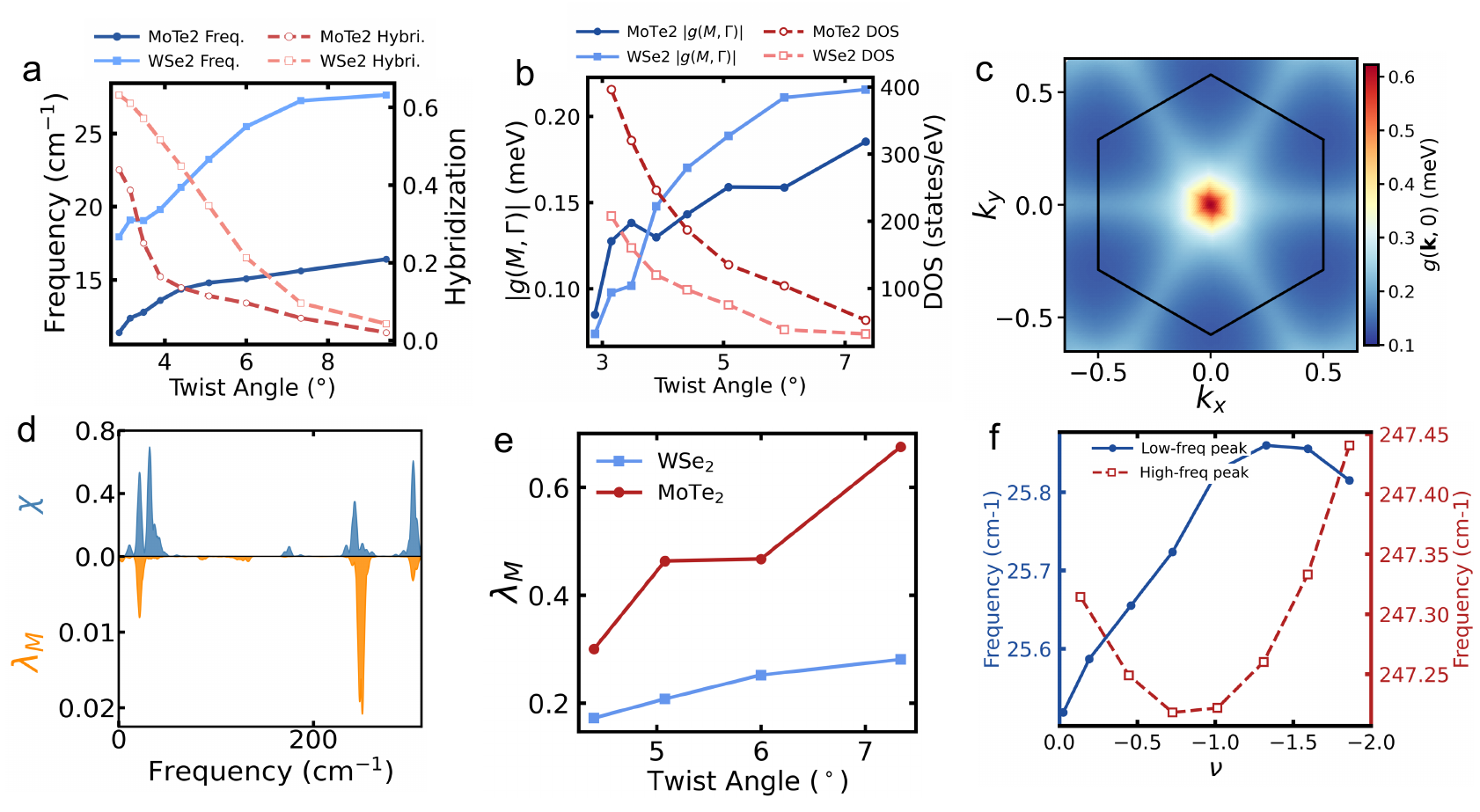}
\caption{ 
\textbf{Twist-angle dependence of the electron--phonon coupling.}
\textbf{a}, Twist-angle dependence of (left) the LBM frequency and (right) the phonon hybridization between the out-of-plane breathing mode and in-plane modes. A hybridization value of 0 denotes a purely out-of-plane breathing mode, while a value of 1 denotes a purely in-plane phonon mode. 
\textbf{b}, Absolute magnitude of the EPC matrix element $g(M,\Gamma)$ of the first band at the $M$ point. 
\textbf{c}, Distribution of the LBM-induced EPC matrix element $g^{\mathrm{LBM}}(\mathbf{k},\Gamma)$ for t-WSe$_2$ at a twist angle of $5.08^\circ$, evaluated on a $100 \times 100$ $\mathbf{k}$-mesh. 
\textbf{d}, Mode-resolved similarity $\chi$ (upper panel) and mode-resolved coupling constant $\lambda_M$ (lower panel). 
\textbf{e}, Total $\lambda_M$ as a function of twist angle for t-WSe$_2$ and t-MoTe$_2$.
f, Measured filling dependence of the low- and high-frequency Raman-active mode positions in t-WSe$_2$ (experiment; cf.\ Fig.~\ref{raman}).
}
\label{TC}
\end{figure*}

We take the $3.15^{\circ}$ and $5.08^{\circ}$ t-MoTe$_2$ structures as representative and examine the impact of all low-frequency phonon branches ($0\text{--}50\text{ cm}^{-1}$) on the electronic band structure. Although the moir\'e superlattice introduces numerous folded phonon modes, the low-frequency spectrum is dominated by modes originating from the $\Gamma$-point of the pristine bilayer: three acoustic modes, two LSMs, and one LBM.
The acoustic modes leave the electronic properties essentially unchanged (Fig.~S7). The LSMs, which strongly influence the electronic properties of twisted bilayer graphene~\cite{wu2018theory,lian2019twisted,koshino2019moir,ochoa2019moire,lu2022low,birkbeck2025quantum}, have a negligible impact here (Fig.~\ref{vibration-pattern}f).
The remaining mode, the LBM, presents a distinct behavior. As shown in Fig.~\ref{vibration-pattern}c, the LBM has a pronounced impact on the electronic structure. 
In the pristine bilayer, the LBM eigenvector is purely out-of-plane, corresponding to a rigid out-of-phase oscillation of the two layers along the $z$ axis. This out-of-plane character is preserved in the moir\'e systems at large twist angles, where the $\Gamma$-point phonon DOS confirms that the LBM displacement remains almost entirely out-of-plane as shown in Figs.~\ref{moire-phonon}e--g. As the twist angle decreases, however, the LBM hybridizes increasingly with in-plane modes. This mode hybridization grows monotonically as the angle is reduced as shown in Fig.~\ref{TC}a, so that the in-plane weight of the eigenvector becomes progressively larger and the small-angle LBM is ultimately dominated by its in-plane component. 
To identify which component most strongly couples to the electronic states, we compute the band structure separately with only the in-plane or only the out-of-plane part of the LBM frozen in (see the Supplementary Material). The electronic structure is far more sensitive to the in-plane component of the LBM, whereas the out-of-plane component has a negligible effect.
Finally, the LBM frequency softens progressively with decreasing twist angle as shown in Fig.~\ref{TC}b. This softening is a direct consequence of the lattice relaxation. After relaxation, the MX/XM domains are pulled to a smaller interlayer spacing, whereas the MM regions are pushed to a larger one. With decreasing twist angle, the MM regions are compressed into smaller, more isolated pockets. Since the LBM is mainly localized in these regions, its restoring force is dominated by the locally softer interlayer potential there, which lowers the average LBM curvature and thus softens the frequency.

We quantify the coupling strength between the LBM and the electronic structure through the matrix element $|g|$, calculated using the finite-displacement method for t-WSe$_2$ and t-MoTe$_2$ over a twist angle range of $3.15^{\circ}$--$7.34^{\circ}$:
\begin{equation}
|g^{j}_{nn}(\mathbf k,\Gamma)| \approx Q_{\mathrm{zpf},j} \left| \frac{\varepsilon_{\mathbf{k},n}(+Q) - \varepsilon_{\mathbf{k},n}(-Q)}{2Q} \right|
\end{equation}
A detailed derivation is provided in the Supplementary Material. Here, $Q_{\mathrm{zpf},j}=\sqrt{\hbar/2\omega_{j}}$ is the zero-point amplitude of the mode, $\omega_{j}$ denotes the phonon frequency, $Q$ represents the normal-mode amplitude (in units of $\mathrm{\AA}\sqrt{\mathrm{amu}}$), and $\varepsilon_{\mathbf{k},n}(\pm Q)$ is the band energy of the distorted structures.
Motivated by the experimental observation of superconductivity at half filling of the first moir\'e band, where the Fermi level lies near the van Hove singularity at the $M$ point of the mBZ, we evaluate the LBM-induced EPC at this momentum. Since the first band is energetically isolated from the remaining bands at $M$, the off-diagonal interband matrix elements are negligible, and the diagonal element $g(M,\Gamma)$ provides a sufficient measure of the coupling strength. 
As shown in Fig.~\ref{TC}b, $g(M,\Gamma)$ decreases monotonically with decreasing twist angle. 
At smaller angles the moir\'e supercell contains more atoms and therefore hosts a larger number of phonon modes, so that the EPC strength is distributed over more modes and the weight carried by an individual LBM is diluted. The electronic DOS shows the opposite trend, increasing toward smaller angles as the moir\'e bands flatten and their bandwidth narrows.

\subsection{Interference selection rule}The origin of this selectivity is a close resemblance between the LBM displacement pattern and the lattice relaxation itself. Because relaxation drastically reconstructs the electronic bands of twisted TMDs such as WSe$_2$ and MoTe$_2$, the electronic structure is exceptionally sensitive to atomic displacements along the relaxation direction, where the deformation potential $\partial V/\partial \mathbf{u}$, the change in the electronic potential $V(\mathbf{r})$ per unit displacement $\mathbf{u}$, is largest.
When the displacement pattern of a phonon mode is closely aligned with this relaxation path, the mode moves the atoms along the very direction in which the electronic potential varies most rapidly. The projection of $\partial V/\partial\mathbf{u}$ onto its eigenvector is then maximal, so the lattice perturbation produced by the mode elicits the largest linear response in the electronic structure. This is precisely the situation of the LBM.
The opposite holds when a mode vibrates along a direction orthogonal to the relaxation path. For rigid-body translations such as the acoustic modes, the uniform displacement leaves the electronic potential unchanged and the derivative itself vanishes. For certain transverse bending modes such as the LSM, the eigenvector is orthogonal to $\partial V/\partial\mathbf{u}$, so the coupling integral cancels. In both cases the EPC matrix element approaches zero, and these modes have a negligible effect on the electronic properties.

We make this geometric resemblance quantitative with a similarity metric $\chi$:
\begin{equation}
\chi = \frac{\mathbf{u}_{\mathrm{relax}} \cdot \mathbf{u}_{\mathrm{frozen}}}{\left\lVert \mathbf{u}_{\mathrm{relax}} \right\rVert \cdot \left\lVert \mathbf{u}_{\mathrm{frozen}} \right\rVert}
\end{equation}
where $\mathbf{u}_{\mathrm{relax}}$ and $\mathbf{u}_{\mathrm{frozen}}$ are the atomic displacement fields of the lattice relaxation and of the specific phonon mode, respectively. Here $\mathbf{u}_{\mathrm{relax}}$ is the positional difference between the relaxed and unrelaxed structures, and $\mathbf{u}_{\mathrm{frozen}}$ is the phonon-induced displacement from the relaxed structure. The metric projects the phonon eigenvector onto the relaxation deformation path: $\chi \approx 1$ indicates a perfect match, whereas $\chi \approx 0$ means the mode is uncorrelated with the relaxation pattern.

As a first qualitative check, we examine whether the modes singled out by $\chi$ are those that most alter the electronic structure. To this end, we survey the low-frequency $(0-50\text{ cm}^{-1})$ phonons of the $3.15^{\circ}/5.08^{\circ}$ structures and analyze the band deformation induced by each mode along a high-symmetry path.
As shown in Fig.~S8, the LBM is mode No.13, which has the lowest frequency and produces a sizable band splitting; seven further modes also exhibit high similarity, and all of them except mode No.73 modify the electronic structure appreciably. At smaller twist angles, the rapidly increasing number of atoms does not multiply the number of high-similarity modes. At a twist angle of $3.15^\circ$, only five phonon modes have high similarity. As shown in Fig.~S9, modes No.19 and No.60 induce the largest modulation of the band structure, with the LBM (No.19) having the lowest frequency and the largest band splitting.

\subsection{Mode-resolved coupling constant}To quantify this correlation, we compute the matrix element
$g(\mathbf{k},\Gamma)$ on a uniform $100\times100$ $\mathbf{k}$-mesh and integrate it to obtain the mode-resolved coupling constant $\lambda_{j}$ of each phonon mode,
\begin{equation}
\lambda_{j} =
\frac{2}{N(E_{\mathrm{F}})\,\Omega_{j}}
\sum_{\mathbf{k}mn}
\left| g^{j}_{mn}(\mathbf{k},\Gamma) \right|^2
\delta(\varepsilon_{\mathbf{k},n}-E_{\mathrm{F}})
\delta(\varepsilon_{\mathbf{k},m}-E_{\mathrm{F}}),
\end{equation}
where $\Omega_{j}$ is the phonon energy at the $\mathbf{q} = \Gamma$ point, and $N(E_{\mathrm{F}})$ is the DOS at the Fermi energy $E_{\mathrm{F}}$. We set the Fermi energy at the $M$ point of the first valence band (filling $\nu = -1$), and calculate the DOS with a broadening parameter of 1~meV. The total EPC constant follows from summing this mode-resolved contribution over all phonon branches, $\lambda = \sum_{j}\lambda_{j}$. Figure~\ref{TC}c shows the distribution of the LBM-induced coupling $g^{\mathrm{LBM}}(\mathbf{k},\Gamma)$ over the mBZ for t-WSe$_2$ at $5.08^{\circ}$. 

The pattern is symmetric: the coupling is strongest at the $\Gamma$ point, remains substantial at $M$, and is consistently weak at $K$.
Figure~\ref{TC}d compares the mode-resolved $\lambda$ with the similarity $\chi$ across all phonon modes; the sum over all modes gives $\lambda = 0.21$ for t-WSe$_2$ at $5.08^\circ$. In the low-frequency regime, $\chi$ displays two peaks, only the first of which aligns with a peak in $\lambda$, identifying it as the LBM.
The second is the highest $\chi$ peak overall; although it couples only weakly to the electronic states at $M$, we show in Fig.~S10 that it dominates the coupling at $\Gamma$. At higher frequencies, $\chi$ exhibits two further prominent peaks, both of which track the corresponding peaks of $\lambda$. This close correspondence between the $\chi$ and $\lambda$ peaks confirms that the similarity metric reliably identifies the phonon modes that contribute most to the EPC. Summing over all phonon branches, we find that the total coupling constant $\lambda$ reaches a pronounced maximum near $7.34^\circ$ and decreases toward smaller twist angles, in contrast to the growing DOS, as shown in Fig.~\ref{TC}e.
The Raman measurements (Fig.~\ref{TC}f) confirm that the two dominant modes, the low-frequency LBM and the high-frequency OBM, are Raman active, and that their peak positions respond strongly to the filling factor.

The high-frequency channel deserves separate comment. The OBM at $\sim$250~cm$^{-1}$ is an optical mode in which the two layers vibrate out of phase, so it retains the interlayer-breathing character that aligns with the reconstruction field. It therefore appears among the high-$\chi$ modes and, consistent with the selection rule, tracks a sizable $\lambda$ peak [Fig.~\ref{TC}d and Fig.~S10]. Unlike the LBM, it lies well above the interlayer manifold and does not soften with twist angle; its large $\chi$ nonetheless makes it the dominant high-frequency contributor to the coupling and the origin of the $250~\mathrm{cm}^{-1}$ Raman peak.

\section{Discussion}
Together, our measurements and calculations identify the moir\'e phonons that couple to the electrons and fix the scale of that coupling. The interference selection rule explains why so few modes matter and why the coupling is strongest at large twist angles, precisely where superconductivity is observed. The rule is not specific to the LBM: several higher-frequency optical branches, including the OBM seen in Raman, also have large $\chi$ and contribute to the total $\lambda$, so the low- and high-frequency modes that dominate the coupling are the very modes resolved in our measurements. For $4.4^\circ$ t-WSe$_2$ the $\Gamma$-point contribution is $\lambda = 0.17$, a lower bound because finite-$q$ phonons are omitted; a finite-$q$ contribution comparable to the $\Gamma$-point term yields a $T_c$ of about $0.3$~K for a Coulomb pseudopotential $\mu^{\ast}\approx0.1$, matching experiment~\cite{guo2025angle}. Any complete theory of large-angle pairing in twisted TMDs must therefore include the electron--phonon coupling.

Pinning down $T_c$ requires the finite-$q$ EPC and the full Eliashberg function, which our framework can supply by extending the frozen-phonon calculation across the moir\'e Brillouin zone. More broadly, the interference selection rule provides a guiding principle for identifying which moir\'e systems favor phonon-driven superconductivity, and applying the same machine-learning framework across twisted TMDs, heterobilayers, and related moir\'e semiconductors would turn that principle into a predictive screening tool. Because phonons and electronic correlations coexist in these systems, a treatment that combines the mode-resolved coupling computed here with the repulsive Kohn--Luttinger channel is a natural next step toward settling the pairing mechanism.

\section*{Acknowledgments}
Y.Z. was supported by the Max Planck partner lab for quantum materials. N.M., C.X., and C.F. acknowledge financial support by the Deutsche Forschungsgemeinschaft (DFG, German Research Foundation) through the Würzburg-Dresden Cluster of Excellence ctd.qmat – Complexity, Topology and Dynamics in Quantum Matter (EXC 2147, project-id 390858490).





\bibliographystyle{apsrev4-2}

\clearpage
\pagebreak
\onecolumngrid
\setcounter{figure}{0}
\renewcommand{\thefigure}{S\arabic{figure}}
\setcounter{equation}{0}
\renewcommand{\theequation}{S\arabic{equation}}
\setcounter{table}{0}
\renewcommand{\thetable}{S\arabic{table}}

\appendix
\section{Methods}

\subsection{Sample fabrication}
Twisted WSe$_2$ devices were fabricated using a standard tear-and-stack dry-transfer method. Atomically thin flakes of WSe$_2$, h-BN, graphite and TaSe$_2$ were exfoliated from bulk crystals onto Si/SiO$_2$ substrates with a 285~nm oxide layer. Graphite flakes with thicknesses of 3--5~nm were used as top and bottom gates, while few-layer TaSe$_2$ flakes with thicknesses of 2--5~nm served as contact electrodes to facilitate hole injection into WSe$_2$ owing to their high work function. Suitable flakes were identified by optical contrast, and the h-BN thicknesses were measured by atomic force microscopy (AFM).

A monolayer WSe$_2$ flake was mechanically cut into two pieces using an AFM tip and assembled into a twisted bilayer inside a N$_2$-filled glovebox, where both oxygen and water levels were kept below 0.01~ppm. The stack was assembled with a polycarbonate stamp, and the twist angle was controlled by a high-precision rotation stage (Thorlabs PRM1Z8) with a typical accuracy of approximately 0.1$^\circ$. The completed heterostructure was transferred onto pre-patterned Ti/Au electrodes with thicknesses of 5~nm and 20~nm, respectively, with the graphite gates and TaSe$_2$ grounding contact aligned to the electrodes. The resulting device was fully encapsulated by h-BN in a dual-gated geometry. Residual polycarbonate was removed by soaking the device in anhydrous chloroform for 5~min, followed by repeated rinsing with acetone and isopropyl alcohol.

\subsection{Raman measurement}

Raman spectra were acquired in backscattering geometry using a home-built confocal microscope integrated with a closed-cycle helium cryostat (attoDRY 2100, base temperature 1.7~K). Excitation was derived from a 532~nm solid-state laser, with the beam focused along the crystallographic $c$-axis onto the dual-gated t-WSez devices through a low-temperature microscope objective (Attocube, 0.8~NA). The same objective served to collect the backscattered signal, which was subsequently coupled into a spectrometer equipped with a 1,800-groove/mm grating and a silicon CCD detector (Princeton Instruments PIXIS: 100B eXcelon). To suppress laser sidebands, a reflective Bragg grating combined with three Bragg notch filters was employed, enabling reliable detection of Raman shifts down to approximately 5~cm$^{-1}$. The incident laser power was set to 1.2~mW.

The Raman data were processed as follows. First, to eliminate interference from fluorescence backgrounds, an Asymmetric Least Squares (ALS) algorithm was applied to each raw spectrum. This method iteratively fits a baseline using a weighted penalized least-squares approach, effectively stripping the background while preserving the peak shapes. The smoothing parameter $\lambda$ was set to $1 \times 10^{5}$, the asymmetry weight $p$ to 0.01, and the number of iterations to 10. Second, cosmic rays were removed by detection using the standard deviation (std) algorithm and correction via median filtering, with a sliding window size of 5 and a detection threshold of 5.0. Finally, the spectra were truncated to the characteristic wavenumber regions of the target Raman peaks, specifically 25--40~cm$^{-1}$ and 245--260~cm$^{-1}$ in this study. Single-peak fitting was performed on each spectrum using a Lorentzian profile via the nonlinear least-squares method with the Levenberg--Marquardt algorithm.

\subsection{Machine-learning force field for phonons}

A second-stage structural relaxation is performed using the Vienna Ab initio Simulation Package (VASP)~\cite{vasp}. We employ Projector Augmented Wave (PAW) potentials, treating six valence electrons for the W, Se, Te, and Mo atoms. The calculations use a plane-wave energy cutoff of 300 eV, and Brillouin-zone sampling is restricted to the $\Gamma$ point. The convergence criteria are set to $1\times10^{-6}$ eV for the electronic self-consistent loop, and the structures are relaxed until the maximum force on each atom drops below 10 meV/$\mathrm{\AA}$. For twisted bilayer MoTe$_2$ (t-MoTe$_2$), we use the dDsC scheme (IVDW=4), whereas for twisted bilayer WSe$_2$ (t-WSe$_2$) we adopt the D2 scheme (IVDW=10).

The training set is prepared by performing fixed-volume relaxations on $3 \times 3 \times 1$ MM, MX, and XM configurations alongside 28 transition states. To diversify the data, 200 structures are generated for each configuration via random perturbations: atomic displacements within $\pm 0.01\mathrm{\AA}$ and lattice strains within $\pm 3\%$. Finally, 20-fs ab initio molecular-dynamics (AIMD) simulations at 300 K are conducted to collect VASP-calculated energies, forces, and virial tensors. This dataset is subsequently employed to train the machine-learning force field (MLFF) model using the DeePMD-kit package~\cite{zhang2018deep}. For transfer learning, we construct a second dataset comprising 118 sets of forces, energies, and virial information.

\end{document}